\documentclass{desyprocA4}
\newcommand\pubdate{\today}
\def\eslt{E_T^{\rm miss}}

\def\esl{\not\!\!{E}}

\def\to{\rightarrow}

\def\bi{\begin{itemize}}
 \def\ei{\end{itemize}}
\def\te{\tilde e}

\def\c1p{C1^\prime}
\def\msq3{\overline{m}_{\tilde{q}}(3)}
\def\ta{\tilde a}
\def\tG{\tilde G}

\def\tl{\tilde l}

\def\ta{\tilde a}

\def\tb{\tilde b}

\def\tst{\tilde t}
\def\ttau{\tilde \tau}
\def\tmu{\tilde \mu}
\def\tg{\tilde g}
\def\tnu{\tilde\nu}
\def\tell{\tilde\ell}
\def\tq{\tilde q}

\def\be{\begin{equation}}  
\def\ee{\end{equation}}  
\def\bea{\begin{eqnarray}}  
\def\eea{\end{eqnarray}}  
\def\tw{\tilde\chi}
\def\twp{\tilde\chi^+}
\def\twm{\tilde\chi^-}
\def\twpm{\tilde\chi^\pm}
\def\tz{\tilde\chi^0}
\newcommand{\alt}{\mbox{$\;\raisebox{-1mm}{$\stackrel{\scriptstyle<}{\scriptstyle\sim}$}\;$}}
\newcommand{\agt}{\mbox{$\;\raisebox{-1mm}{$\stackrel{\scriptstyle>}{\scriptstyle\sim}$}\;$}}

\def\beq{\begin{equation}}
\def\eeq#1{\label{#1}\end{equation}}
\def\eeqn{\end{equation}}

\newenvironment{Eqnarray}%
   {\arraycolsep 0.14em\begin{eqnarray}}{\end{eqnarray}}
\def\beqa{\begin{Eqnarray}}
\def\eeqa#1{\label{#1}\end{Eqnarray}}
\def\eeqan{\end{Eqnarray}}

\newcommand{\ra}            {\ensuremath{ \rightarrow     }}

\begin{document}
\title{Post LHC7 SUSY benchmark points for ILC physics}

\author{{\slshape Howard Baer$^1$, Jenny List$^2$}\\[1ex]
$^1$University of Oklahoma, Norman, OK 73019, USA\\
$^2$DESY, Notkestra{\ss}e 85, 22607 Hamburg, Germany\\
}

\confID{4980}  
\desyproc{LC-REP-2012-063, DESY 12-086}
\doi  

\maketitle

\pubdate

\begin{abstract}
We re-evaluate prospects for supersymmetry at the proposed 
International Linear $e^+e^-$ Collider (ILC) in light of the first
year of serious data taking at LHC with $\sqrt{s}=7$ TeV and $\sim 5$ fb$^{-1}$ of 
$pp$ collisions (LHC7).
Strong new limits from LHC SUSY searches, along with a hint of a Higgs boson
signal around $m_h\sim 125$ GeV, suggest a paradigm shift from previously popular models
to ones with new and compelling signatures. We present a variety of new ILC benchmark
models, including: natural SUSY, hidden SUSY, NUHM2 with low $m_A$, non-universal gaugino mass
(NUGM) model, pMSSM, Kallosh-Linde model, Br\"ummer-Buchm\"uller model, 
normal scalar mass hierarchy (NMH) plus one surviving case from mSUGRA/CMSSM in the far
focus point region. 
While all these models at present elude the latest LHC limits, 
they do offer intriguing case study possibilities for ILC operating at $\sqrt{s}\sim 0.25-1$ TeV,
and present a view of some of the diverse SUSY phenomena which might be expected at both
LHC and ILC in the post LHC7 era.
\end{abstract}

\section{Introduction}

\subsection{Motivation}

Supersymmetry (SUSY) is a quantum spacetime symmetry which predicts a correspondence between
bosonic and fermionic fields~\cite{Wess:1973kz,Salam:1974jj,Salam:1974ig,Baer:2006rs}. 
Supersymmetry is particularly appealing for theories of
particle physics in that it reduces scalar field quadratic divergences to merely logarithmic. 
This fact allows for an elegant solution to the notorious gauge hierarchy problem, rendering
the weak scale stable against quantum corrections and allowing for stable extrapolations of the
Standard Model (SM) into the far ultraviolet ($E\gg M_{weak}$) regime~\cite{Witten:1981nf,Kaul:1981wp}. 
Thus, SUSY provides an 
avenue for connecting the Standard Model to ideas of grand unification (GUTs) and/or string theory,
and provides a route to unification with gravity via local SUSY, or supergravity 
theories~\cite{Ferrara:1976kg,Cremmer:1982en,Nilles:1983ge}. 

While models of weak scale supersymmetry are theoretically compelling, we note here 
that a variety of indirect evidence from experiment has emerged which provides
support for the idea of weak scale SUSY:
\begin{itemize}
\item {\it Gauge coupling unification:} The values of the three SM gauge couplings,
measured at energy scale $Q\simeq M_Z$ at the CERN LEP collider, 
when extrapolated to high energy scales via renormalization group (RG) running in the
Minimal Supersymmetric Standard Model (MSSM)~\cite{Dimopoulos:1981yj}, very nearly meet at a point 
around $Q\simeq 2\times 10^{16}$ GeV~\cite{Amaldi:1991cn,Langacker:1991an,Ellis:1990wk}. 
Unification of gauge couplings is predicted by many grand unified theories (GUTs) and
string theories. Gauge coupling unification is violated by numerous standard deviations 
under SM RG running.
\item {\it Precision electroweak measurements:} Fits of precision electroweak
observables (EWPO)  to SUSY model predictions find accord provided there exists a rather heavy SUSY 
particle mass spectrum~\cite{Heinemeyer:2006px}. 
Meanwhile, models such as minimal technicolor are highly stressed if not ruled
out by EWPO.
\item {\it Top quark mass and electroweak symmetry breaking:} The electroweak scalar potential is highly
constrained in SUSY theories compared to the SM, and it is not immediately clear if electroweak symmetry
can be properly broken, yielding the required vector boson and fermion masses while leaving the 
photon massless. 
In top-down theories, the soft breaking Higgs mass $m_{H_u}^2$ is driven to negative values
by the large top quark Yukawa coupling, triggering an appropriate breakdown of EW symmetry, 
provided that the top quark mass $m_t\sim 150-200$ GeV~\cite{Ibanez:1982fr}. 
The latest measurements find $m_t=173.2\pm 0.9$ GeV~\cite{Lancaster:2011wr}.
\item {\it Higgs mass:}  In the SM, the physical Higgs scalar mass $m_{H_{SM}}>115$ GeV due to LEP2 and LHC searches,
and it is lighter than $\sim 800$ GeV~\cite{Reina:2005ae} from unitarity constraints~\cite{Lee:1977eg}.
In the MSSM, typically $m_A\gg m_h$ so that $h$ is SM-like. In this case, $m_h>115$ GeV as in the SM case, but
also $m_h\alt 135$ GeV due to its more constrained mass calculation including radiative corrections~\cite{Djouadi:2008gy}.
The latest data from the CERN LHC and Fermilab Tevatron is consistent with 
$115\ {\rm GeV}<m_h<127\ {\rm GeV}$ with
a $(2-3)\sigma$ evidence for $m_h\simeq 125$ GeV~\cite{bib:CMS_SMHiggs,bib:ATLAS_SMHiggs,bib:Tevatron_SMHiggs}, 
squarely in the narrow SUSY window of consistency. 
\item {\it Dark matter:} While none of the SM particles have the right properties to constitute
cold dark matter in the universe, SUSY theories offer several candidates~\cite{Steffen:2008qp}. 
These include the neutralino (a WIMP candidate), the gravitino or a singlet sneutrino. 
In SUSY theories where the strong $CP$
problem is solved via the Peccei-Quinn mechanism, there is the added possibility of mixed 
1. axion-neutralino~\cite{Choi:2008zq,Baer:2011hx,Baer:2011uz},
2. axion-axino~\cite{Rajagopal:1990yx,Covi:2001nw,Baer:2009ms} or 
3. axion-gravitino cold dark matter.
\item {\it Baryogenesis:} The measured baryon to photon ratio $\eta\sim 10^{-10}$ is not possible to explain in
the SM. In SUSY theories, three prominent possibilities include 1. electroweak baryogenesis (now nearly excluded
by limits on $m_{\tst_1}$ and $m_h$~\cite{Curtin:2012aa}), 
2. thermal and non-thermal leptogenesis~\cite{Buchmuller:2005eh}, and 
3. Affleck-Dine baryo- or leptogenesis~\cite{Affleck:1984fy,Dine:1995kz}.
\end{itemize}

\subsection{Some problems for SUSY models}

While the above laundry list is certainly compelling for the existence of weak scale SUSY in nature,
we are faced with the fact that at present there is no evidence for direct superparticle production at high energy colliders, 
especially at the CERN Large Hadron Collider (LHC). The ATLAS and CMS experiments have accumulated $\sim 5$ fb$^{-1}$
of integrated luminosity from $pp$ collisions at $\sqrt{s}=7$ TeV in 2011 (LHC7), and they anticipate collecting
$\sim 15$ fb$^{-1}$ at $\sqrt{s}=8$ TeV in 2012 (LHC8). 
Recent analyses by the CMS experiment~\cite{bib:CMS_razor} using 4.4 fb$^{-1}$ of data have now excluded $m_{\tg}\alt 1400$ GeV 
in the mSUGRA (also known as CMSSM) model,
for the case of $m_{\tq}\simeq m_{\tg}$, while values of $m_{\tg}\alt 800$ GeV are excluded in the case where $m_{\tq}\gg m_{\tg}$.
Indeed, fits of the mSUGRA model as recently as 2010~\cite{Buchmueller:2010ai} to a variety of observables including EWPO, 
$(g-2)_\mu$, $B$-meson decay branching fractions and neutralino cold dark matter density 
predicted SUSY to lie exactly in this excluded range.
In addition, if the light SUSY Higgs boson turns out to have $m_h\simeq 125$ GeV, 
then the minimal versions of gauge-mediated and anomaly-mediated SUSY breaking models will 
likely be ruled out~\cite{Arbey:2011ab}, 
since it is difficult to obtain such large values of $m_h$ in these models unless the 
sparticle mass spectra exist with a lightest MSSM particle with mass greater than
about 5 TeV~\cite{Baer:2012uy}.

While the above results may seem disconcerting, at the same time they were not unanticipated by many theorists.
Whereas SUSY theories solve a host of problems as mentioned above, they also bring with them considerable
phenomenological baggage~\cite{Dienes:1997nq}. Some of these SUSY problems include the following:
\begin{itemize}
\item The SUSY flavor problem~\cite{Gabbiani:1996hi}:
In SUSY models based on gravity-mediation, it is generally expected that
large flavor-violating terms will occur in the Lagrangian~\cite{Kaplunovsky:1993rd}, giving rise to large contributions to the
kaon mass difference, and flavor violating decays such as $b\to s\gamma$ or $\mu\to e\gamma$.
Solutions to the SUSY flavor problem include 
1. degeneracy of matter scalar masses, in which case a SUSY GIM mechanism suppresses flavor violation~\cite{Dimopoulos:1981zb}, 
2. alignment of squark and quark mass matrices~\cite{Nir:1993mx}, or 
3. decoupling mainly of first/second generation scalars 
($m_{\tq,\tell}\agt 5-50$ TeV)~\cite{Dine:1990jd,Cohen:1996vb,ArkaniHamed:1997ab}.\footnote{
Some degree of alignment or degeneracy would still be needed for the lower portion of this 
mass range.} 
Indeed, the SUSY flavor problem provided strong impetus
for the development of GMSB and AMSB models, where universality of scalars with the same quantum numbers is automatically expected.

\item The SUSY $CP$ problem: In this case, it is expected in gravity mediation that $CP$-violating phases in the
soft SUSY breaking terms and perhaps $\mu$ parameter will give rise to large electron and neutron (and other) electric dipole moments
(EDMs).
Solutions include dialing the $CP$-violating phases to zero, or decoupling. 
Models such as GMSB and AMSB are again not expected to have complex, CP-violating soft terms.

\item Proton decay in SUSY GUT theories: In SUSY GUT theories, the proton is expected to decay
to $K^+\bar{\nu}$ via colored Higgsino $\tilde h$ exchange. The lifetime is expected
to occur at levels below experimental limits~\cite{Murayama:2001ur}. 
Since $\tau_p\sim m_p^5/m_{\tilde h}^2m_{\tq}^2$, large squark masses can again suppress proton decay.

\item The gravitino problem~\cite{Weinberg:1982zq}: 
In models of gravity-mediation, the superhiggs mechanism generates SUSY breaking
by giving the gravitino a mass $m_{3/2}$. The gravitino mass sets the scale for the visible sector soft breaking terms,
and so one expects sparticle masses of order $m_{3/2}$. However, thermal production of gravitinos in the early universe
can lead to either 1. an overproduction of dark matter (here, the gravitinos would decay to the stable LSP, or even be the LSP),  
or 2. late-time decays of gravitinos at time scales $\agt 1$~s after the Big Bang would lead to dissolution
of the light nuclei built up during Big Bang nucleosynthesis (BBN). Solutions to the gravitino problem include
1. a rather low re-heat temperature $T_R\alt 10^5$ GeV after inflation so that thermal gravitino production is suppressed~\cite{Khlopov:1984pf} 
(but such low $T_R$ values conflict with some baryogenesis mechanisms such as leptogenesis, which 
seems to require $T_R\agt 10^9$ GeV), 
2. a rather light gravitino with $m_{3/2}\ll 1$ GeV, which enhances the goldstino coupling, or
3. a rather heavy gravitino $m_{3/2}\agt 5$ TeV, which lowers the gravitino lifetime so that $\tau_{3/2}\alt 1$ sec, and
gravitinos decay before BBN~\cite{Kawasaki:2008qe}.
\end{itemize}

While some proposed solutions solve individual problems listed above ({\it e.g.} alignment for the SUSY flavor problem,
low $T_R$ for the gravitino problem, small phases for the SUSY $CP$ problem), there is one solution-- 
{\it decoupling of first/second generation matter scalars}-- which goes a long way to solving all 
four.\footnote{In gravity mediation, it is expected that the gravitino mass
$m_{3/2}$ sets the mass scale for the heaviest of the scalars; in this case, multi-TeV scalar masses would proceed from
a multi-TeV gravitino mass.} But what of fine-tuning constraints in SUSY models, which seemingly
require sparticle masses near the weak scale~\cite{Anderson:1994tr}?

\subsection{Fine-tuning in supersymmetric models}

The connection between the SUSY breaking scale and the magnitude of the weak scale can be 
understood most directly by minimization of the scalar potential in the MSSM 
to determine the magnitude of the electroweak vacuum expectation values.
The scalar potential gains contributions from three sectors:
\begin{equation}
V_{SUSY}=V_F+V_D+V_{soft} ,
\end{equation}
and with 50 field ``directions'' in the MSSM, the scalar potential is rather daunting. Under rather mild conditions, 
charge and color breaking minima can be avoided, so that instead we just minimize in the neutral/non-colored 
scalar field directions.
A well-defined local minimum can be found where the vacuum expectation values of the real parts of the neutral Higgs fields are given by
$\langle h_u^0\rangle\equiv v_u$ and $\langle h_d^0\rangle\equiv v_d$ with $\tan\beta\equiv v_u/v_d$. The $Z$ boson acquires a 
mass $M_Z^2=\frac{g^2+g^{\prime 2}}{2}\left(v_u^2+v_d^2\right)$. Including radiative corrections, 
the scalar potential minimization condition is then written as
\begin{equation}
 \frac{1}{2}M_Z^2
=\frac{(m_{H_d}^2+\Sigma_d)-(m_{H_u}^2+\Sigma_u)\tan^2\beta}{(\tan^2\beta
  -1)} - \mu^2\;. 
\label{eq:zmass}
\end{equation}
Here, $\Sigma_u$ and $\Sigma_d$ arise from radiative corrections~\cite{Arnowitt:1992qp},
and are given in the 1-loop approximation to the Higgs effective potential by
$$\Sigma_{u,d}= \frac{1}{v_{u,d}}\frac{\partial \Delta V}{\partial H_{u,d}},$$ 
where $\Delta V$ is the one-loop correction to the tree-level potential,
and the derivatives are evaluated at the physical vacuum.
  
It is then reasonable to say that the theory yields a natural value of $M_Z$
if the individual terms on the right hand side of Eq.~(\ref{eq:zmass})
are comparable in magnitude so that the observed value of $M_Z$ is
obtained without resorting to large cancellations. Indeed this is why
$|\mu|$ has been suggested as a measure of naturalness~\cite{Chan:1997bi},
with theories where $\mu^2 \alt M_Z^2$ being the ``most natural''. 
This relationship must be accepted with some latitude, since values of 
$\mu^2 \alt$ (100~GeV)$^2$ are phenomenologically excluded. 
Here, we will adopt $|\mu |<\Lambda_{NS}$, where $\Lambda_{NS}\sim M_Z$, but might be as high
as $\sim 200$ GeV.
Of course, there is nothing special about the magnitude of $\mu$, so that the same  considerations
apply equally to all the terms in Eq'n~\ref{eq:zmass}, including those involving the radiative corrections. 
Naturalness thus requires that each 
{\it individual} term in (\ref{eq:zmass}) be $\alt \Lambda_{NS}$.

The largest contributions to $\Sigma_{u,d}$ in Eq.~(\ref{eq:zmass})
arise from superpotential  Yukawa interactions of third generation
squarks involving the top quark Yukawa coupling. 
The order of magnitude of these contributions is given by
$$\Sigma_u \sim \frac{3f_t^2}{16\pi^2}\times m_{\tst_i}^2 \left(\ln
(m_{\tst_i^2}/Q^2) -1\right)\;,$$ 
and so grows quadratically with the top squark masses. 
Clearly, the top squark (and by $SU(2)$ gauge symmetry, also $\tb_L$) masses must then be
bounded from above by the naturalness conditions. In Ref.~\cite{King:2012is}, it
has been shown that requiring  $\Sigma_u \alt \frac{1}{2}M_Z^2$ leads to
$m_{\tst_i}\alt 500$~GeV. Scaling this up to allow $\mu$ values up to 150-200~GeV
leads to a corresponding bound $m_{\tst_i} \alt 1-1.5$~TeV.
In other words, from this perspective, models with $\mu \alt 200$~GeV and top squarks at the
TeV scale or below are preferred by naturalness. 
It is also worth remarking that since 
\begin{equation}
m_A^2\simeq 2\mu^2 + m_{H_u}^2 + m_{H_d}^2 + \Sigma_u + \Sigma_d\;,
\label{eq:Amass}
\end{equation}
for moderate to large values of $\tan\beta$, 
the heavier Higgs scalars can naturally be at the several-TeV scale because of the appearance of
$\tan^2\beta-1$ in the denominator of Eq.~(\ref{eq:zmass}). 
Notice, however, that the bound of $\Lambda_{NS}^2$ on each term in
Eq.~(\ref{eq:zmass}) translates to an upper bound $m_A \alt\Lambda_{NS} \tan\beta$. 

There will also be
corresponding constraints on other sparticles such as electro-weak
charginos and neutralinos that directly couple to the Higgs sector, but
since these couplings are smaller than $f_t$ and there are no color factors, 
the constraints will be correspondingly weaker. 
Sparticles such as first and second generation squarks and sleptons that have no
direct/significant couplings to the Higgs sector are constrained only via two-loop
effects and can easily be in the 10-50 TeV range. 
An important exception would be the gluino, since radiative corrections to the top squark mass
are proportional to $m_{\tg}$~\cite{Brust:2011tb}.
Using $\delta m_{\tq}^2 \sim \frac{2g_s^2}{3\pi^2}m_{\tg}^2\times log$ and setting logs to be 
order unity, we expect that $m_{\tg} \alt 3m_{\tq}$. 
For top squarks to remain in the $\sim 1.5$~TeV range, the gluino must be lighter than 3-4 TeV. 
In models with electroweak gaugino mass unification, electroweak-inos would then
automatically not destroy naturalness.

To summarize, naturalness considerations suggest that SUSY models should give rise 
to a mass spectrum characterized by
\bi
\item $|\mu |\alt \Lambda_{NS}\sim 200$~GeV, 
\item third generation squarks $m_{\tst_{L,R}},\ m_{\tb_L}\alt 1.5$~TeV,
\item $m_{\tg} \alt 3-4$~TeV and SSB electroweak-ino masses smaller than 1-2~TeV
\item $m_{\tq_{1,2}},\ m_{\tell_{1,2}}\sim 10-50$ TeV.
\ei
The latter weak constraint on first/second generation matter scalars allows for a decoupling
solution to the SUSY flavor, $CP$, $p$-decay and (indirectly) gravitino problems.
SUSY models with the above generic spectra have been dubbed 
``natural SUSY''~\cite{Papucci:2011wy}.\footnote{For earlier related work, see
Ref's~\cite{Dimopoulos:1995mi,Kitano:2005wc,Kitano:2006gv,Kitano:2006ws,Cheung:2005pv,Baer:2011ec}.}
This spectra is closely related to {\it effective SUSY}\cite{Cohen:1996vb}, but with the additional 
requirement that $|\mu |\alt 150-200$ GeV 
which would likely give rise to a higgsino-like lightest neutralino $\tz_1$.
In contrast, models such as mSUGRA with rather heavy top squarks are expected to be highly fine-tuned, 
even when $\mu$ is small as in the hyperbolic branch/focus point (HB/FP) region.

The remainder of this report is geared towards presenting a new set of supersymmetry benchmark
models suitable for ILC investigations, while maintaining consistency with the latest 
indirect and direct constraints on supersymmetric models, especially taking into 
account what has been learned from recent LHC searches. In Sec.~\ref{sec:constraints}, we
briefly summarize current indirect constraints on SUSY models, and also discuss 
the current status of SUSY dark matter.
In Sec.~\ref{sec:lhc}, we present a summary of the most recent results from LHC searches for SUSY and
Higgs bosons. In Sec.~\ref{sec:BMs}, we present a variety of new post LHC7 benchmark points
for ILC studies. These new benchmarks reflect a movement away from previous studies within the
mSUGRA/CMSSM model. Some models have been selected due to their theoretical motivation
({\it e.g.} natural SUSY and its relatives), while others have been selected for their diversity of
phenomenology which may be expected at ILC.
In Sec.~\ref{sec:conclude}, we present a brief summary and outlook for physics 
prospects at the ILC.

\section{Indirect constraints on SUSY models}
\label{sec:constraints}

In this section, we review briefly indirect constraints on SUSY models  from muon $g-2$
measurements, rare $B$-decay branching fractions along with an updated discussion of the role of
dark matter in SUSY models.

\subsection{$(g-2)_\mu$ status}

The magnetic moment of the muon $a_\mu\equiv\frac{(g-2)_\mu}{2}$ 
was measured by the Muon $g-2$ Collaboration~\cite{Bennett:2006fi} 
and has been found to give a $3.6\sigma$ discrepancy with SM calculations based on $e^+e^-$ data~\cite{Davier:2010nc}:
$\Delta a_\mu =a_\mu^{meas}-a_\mu^{SM}[e^+e^-]=(28.7\pm 8.0)\times 10^{-10}$. When $\tau$-decay data are used
to estimate the hadronic vacuum polarization contribution rather than low energy $e^+e^-$ annihilation 
data, the discrepancy reduces to $2.4\sigma$ , corrensponding to 
$\Delta a_\mu =a_\mu^{meas}-a_\mu^{SM}[\tau]=(19.5\pm 8.3)\times 10^{-10}$.

The SUSY contribution to the muon magnetic moment is\cite{Moroi:1995yh}
$\Delta a_\mu^{SUSY}\sim \frac{m_\mu^2\mu M_i\tan\beta}{m_{SUSY}^4}$ where $i=1,2$ stands for
electroweak gaugino masses and $m_{SUSY}$ is the characteristic sparticle mass circulating in the
muon-muon-photon vertex correction: here, $m_{\tmu_{L,R}}$, $m_{\tnu_\mu}$, $m_{\twp_i}$ and $m_{\tz_j}$. 
Attempts to explain the muon $g-2$ anomaly using supersymmetry usually invoke sparticle mass 
spectra with relatively light smuons and/or large $\tan\beta$ 
(see {\it e.g.} Ref.~\cite{Feng:2001tr}). 
Some SUSY models where $m_{\tmu_{L,R}}$ is correlated with squark masses (such as mSUGRA) 
are now highly stressed to explain the $(g-2)_\mu$ anomaly. In addition, since naturalness favors a low value
of $|\mu |$, tension again arises between a large contribution to $\Delta a_\mu^{SUSY}$ and naturalness conditions.
These tensions motivate scenarios with non-universal scalar masses. Of the benchmark scenarios discussed in the 
following, some feature light smuons which raise $(g-2)_\mu$  to its experimental value, while others are 
compatible with the Standard Model prediction.

\subsection{$b\to s\gamma$}

The combination of several measurements of the $b\to s\gamma $ branching fraction finds that 
$BF(b\to s\gamma )=(3.55\pm 0.26)\times 10^{-4}$~\cite{Asner:2010qj}.
This is somewhat higher than the SM prediction~\cite{Misiak:2006zs} of 
$BF^{SM}(b\to s\gamma )=(3.15\pm 0.23)\times 10^{-4}$. SUSY contributions to the
$b\to s\gamma$ decay rate come mainly from chargino-top-squark loops and
loops containing charged Higgs bosons, and so are large when these particles are light 
and when $\tan\beta$ is large~\cite{Baer:1996kv}.

\subsection{$B_s\to \mu^+\mu^-$}

The decay $B_s\to\mu^+\mu^-$ occurs in the SM at a calculated branching ratio value of 
$(3.2\pm 0.2)\times 10^{-9}$. 
The CMS experiment~\cite{Chatrchyan:2011kr} has provided an 
upper limit on this branching fraction of $BF(B_s\to\mu^+\mu^- )<1.9\times 10^{-8}$
at 95\% CL. The CDF experiment~\cite{Aaltonen:2011fi} claims a signal in this channel at 
$BF(B_s\to\mu^+\mu^- )=(1.8\pm 1.0)\times 10^{-8}$ at 95\% CL, which is in some discord
with the CMS result.
Finally, the LHCb experiment has reported a strong new bound of
$BF(B_s\to\mu^+\mu^- )<4.5\times 10^{-9}$\cite{Aaij:2012ac}.
In supersymmetric models, this flavor-changing decay occurs 
through pseudoscalar Higgs $A$ exchange~\cite{Babu:1999hn,Mizukoshi:2002gs}, 
and the contribution to the branching fraction from SUSY is proportional
to $\frac{\tan^6\beta}{m_A^4}$. 

\subsection{$B_u\to\tau^+\nu_\tau$}

The branching fraction for $B_u\to\tau^+\nu_\tau$ decay is calculated~\cite{Eriksson:2008cx} in the SM to be
$BF(B_u\to\tau^+\nu_\tau )=(1.10\pm 0.29)\times 10^{-4}$. This is to be compared to the value from
the Heavy Flavor Averaging group~\cite{Barberio:2008fa}, which finds a measured value of 
$BF(B_u\to\tau^+\nu_\tau )=(1.41\pm 0.43)\times 10^{-4}$, in agreement with the SM prediction, but leaving room for additional
contributions.
The main contribution from SUSY comes from tree-level charged Higgs exchange, 
and is large at large $\tan\beta$ and low $m_{H^+}$.

\subsection{Dark matter}

During the past several decades, a very compelling and simple scenario has
emerged to explain the presence of dark matter in the universe with an abundance roughly
five times that of baryonic matter. The WIMP miracle scenario posits that 
weakly interacting massive particles would be in thermal equilibrium with the cosmic
plasma at very high temperatures $T\agt m_{\mathrm WIMP}$. As the universe expands and cools, 
the WIMP particles would freeze out of thermal equilibrium, locking in a relic abundance
that depends inversely on the thermally-averaged WIMP (co)-annihilation 
cross section~\cite{Lee:1977ua}.
The WIMP ``miracle'' occurs in that a weak strength annihilation cross section gives
roughly the measured relic abundance provided the WIMP mass is of the order of the 
weak scale~\cite{Baltz:2006fm}.
The lightest neutralino of SUSY models has been touted as a 
protypical WIMP candidate~\cite{Goldberg:1983nd,Ellis:1983ew,Jungman:1995df}.

While the WIMP miracle scenario is both simple and engaging, it is now clear that
it suffers from several problems in the case of SUSY theories. 
The first of these is that in general SUSY theories where the lightest neutralino
plays the role of a thermally produced WIMP, the calculated relic abundance $\Omega_{\chi}h^2$
is in fact typically two-to-four orders of magnitude larger than the measured abundance
$\Omega_{CDM}^{meas}h^2\sim 0.11$ in the case of a bino-like neutralino, and one-to-two
orders of magnitude lower than measurements in the case of wino- or higgsino-like
neutralinos~\cite{Baer:2010wm}. In fact, rather strong co-annihilation, resonance annihilation or
mixed bino-higgsino or mixed wino-bino annihilation is needed to obtain the measured
dark matter abundance. Each of these scenarios typically requires considerable large 
fine-tuning of parameters to gain the measured dark matter abundance~\cite{Baer:2009vr}. 
The case where neutralinos naturally give the measured CDM
abundance is when one has a bino-like neutralino annihilating via slepton exchange
with slepton masses in the 50-70 GeV range: such mass values were long ago ruled out by slepton
searches at LEP2.

The second problem with the SUSY WIMP miracle scenario is that it neglects the gravitino, which
is an essential component of theories based on supergravity. Gravitinos can be produced 
thermally at high rates at high re-heat temperatures $T_R$ after inflation. 
If $m_{\tG}>m_{LSP}$, then gravitino decays into a stable LSP can overproduce 
dark matter for $T_R\agt 10^{10}$ GeV. 
Even at much lower $T_R\sim 10^5-10^{10}$ GeV, 
thermal production of gravitinos followed by late decays 
(since gravitino decays are suppressed by the Planck scale) tend to dissociate light nuclei
produced in the early universe, thus destroying the successful picture of 
Big Bang nucleosynthesis~\cite{Kawasaki:2008qe}.

The third problem is that the SUSY WIMP scenario neglects at least two very compelling 
new physics effects that would have a strong influence on dark matter production in the 
early universe. 
\bi
\item The first of these is that string theory seems to require 
the presence of at least one light ($\sim 10-100$ TeV) moduli field~\cite{Acharya:2010af}. 
The moduli can be produced at large rates in the early universe and decay 
at times $\sim 10^{-1}-10^5$ sec after the Big Bang. 
Depending on their branching fractions, they could
either feed additional LSPs into the cosmic plasma~\cite{Moroi:1999zb}, 
or decay mainly to SM particles, thus diluting all relics present at the time of 
decay~\cite{Gelmini:2006pq}.
\item The second neglected effect is the strong $CP$ problem, which is deeply routed in QCD
phenomenology~\cite{Peccei:2006as}. 
After more than three decades, the most compelling solution to the strong $CP$
problem is the hypothesis of a Peccei-Quinn axial symmetry whose breaking gives rise to
axion particles with mass $\sim 10^{-6}-10^{-9}$ eV~\cite{Kim:2008hd}. The axions can be produced non-thermally
via coherent oscillations~\cite{Abbott:1982af,Preskill:1982cy,Dine:1982ah}, 
and also would constitute a portion of the dark matter. 
In SUSY theories, the axions are accompanied by $R$-odd spin-${1\over 2}$ axinos $\ta$ and
$R$-even spin-0 saxions $s$~\cite{Nilles:1981py}. 
Thermal production of axinos and non-thermal production of saxions can either
feed more dark matter particles into the cosmic plasma, or inject additional entropy, thus diluting
all relics present at the time of decay. Theoretical predictions for the relic abundance of
dark matter in these scenarios are available but very model-dependent. In the case of 
mixed axion-neutralino dark matter, it is usually very difficult to lower a standard
overabundance of neutralinos, but it is also very easy to bolster a standard underabundance~\cite{Baer:2011uz}.
This latter case may lead one to consider SUSY models with a standard underabundance
of wino-like or higgsino-like neutralinos as perhaps the more compelling possibility for
CDM. In the case of mixed axion-neutralino CDM, it can be very model-dependent
whether the axion or the neutralino dominates the DM abundance, and  cases where
there is a comparable admixture of both are possible. 
\ei
The upshot for ILC or LHC physics is that one shouldn't take dark matter abundance constraints on
SUSY theories too seriously at this point in time.

\subsubsection{Status of WIMP dark matter searches}

As of spring 2012, a variety of direct and indirect WIMP dark matter detection searches
are ongoing. Several experiments -- DAMA/Libra, CoGent and Cresst -- claim excess signal
rates beyond expected backgrounds. These various excesses can be interpreted in terms of
a several GeV WIMP particle, although the three results seem at first sight inconsistent 
with each other. It is also possible that muon or nuclear decay induced 
neutron backgrounds -- which are very difficult to estimate -- contribute to the 
excesses. Numerous theoretical and experimental analyses are ongoing to sort 
the situation out. A  WIMP particle of a few GeV seems hard to accommodate in SUSY theories. 

There also exists claims for measured positron excesses in cosmic rays above expected backgrounds
by the Pamela collaboration~\cite{Adriani:2008zr} and claims for an electron 
excess by the Fermi-LAT group~\cite{Ackermann:2010ij}. 
While these claims can be understood in terms of very massive
WIMPs of order hundreds of GeV, it is unclear at present whether the positrons arise from exotic
astrophysical sources~\cite{Profumo:2008ms} or simply from rare mis-identification of cosmic protons.

A variety of other direct WIMP search experiments have probed deeply into WIMP-model
parameter space, with no apparent excesses above SM background. At this time, the best limits come
from the Xenon-100 experiment~\cite{Aprile:2011hi}, 
which excludes WIMP-proton scattering cross sections of
$\sigma (\chi p)\agt 10^{-8}$ pb at 90\%CL for $m_{\mathrm WIMP}\sim 100$ GeV.
The Xenon-100, LUX and CDMS experiments seem poised decisively to probe the expected parameter space
of mixed bino-higgsino dark matter~\cite{Baer:2006te,Feng:2010ef} (as occurs for instance in focus point SUSY of the mSUGRA model)
in the next round of data taking.

\subsubsection{Gravitino dark matter}

It is possible in SUSY theories that gravitinos are the lightest SUSY particle, 
and could fill the role of dark matter. In gravity-mediation, the gravitino is expected 
to have mass of order the weak scale. In this case, late decays of thermally produced
neutralinos into gravitinos are often in conflict with BBN constraints. If the gravitinos are
much lighter, well below the GeV scale, then their goldstino coupling is enhanced
and BBN constraints can be evaded. This scenario tends to occur for instance in
gauge-mediated SUSY theories. The simplest GMSB scenarios now appear in
conflict with Higgs mass results if indeed LHC is seeing $m_h$ at $\sim 125$~GeV~\cite{Arbey:2011ab,Baer:2012uy}. We will, however, present an example of a non-minimal GMSB 
model which is compatible with a Higgs mass of $\sim 125$~GeV.

\section{LHC results}
\label{sec:lhc}

In this section, we present a very brief summary of the status of 
LHC searches for SUSY Higgs bosons and for SUSY particles as of April 2012.

\subsection{Impact of Higgs searches}

\subsubsection{SM-like Higgs scalar}

The ATLAS and CMS experiments reported on search results for a 
SM-like neutral Higgs scalar $H_{SM}$ in March 2012 based on about 5 fb$^{-1}$
of data at $\sqrt{s}=7$ TeV~\cite{bib:CMS_SMHiggs,bib:ATLAS_SMHiggs}. 
Their analyses exclude a SM-like Higgs boson over 
the mass range $127<m_{H_{SM}}<600$ GeV. Combining this
range with a fit of precision electroweak data to SM predictions then 
allows a SM-like Higgs boson to live in the narrow mass range of 
$115\ {\rm GeV}<m_{H_{SM}}<127$ GeV. In fact, ATLAS reports an excess of 
events at $3.5\sigma$ level in the $\gamma\gamma$, $WW^*$ and $ZZ^*$ channels which is 
consistent with $m_{H_{SM}}\sim 126$ GeV.
A similar excess is reported by CMS at $3.1\sigma$ at $m_{H_{SM}}\sim 124$ GeV, 
along with an  excess of $4\ell$ events at $\sim 120$ GeV.
These excesses are also corroborated by recent reports from CDF and D0
at the Fermilab Tevatron of excess events over the mass range 115-130 GeV~\cite{bib:Tevatron_SMHiggs}.
Upcoming data from the 2012 LHC run at $\sqrt{s}=8$ TeV should validate
or exclude a Higgs signal in the 115-127 GeV range.

\subsubsection{Non-standard Higgs bosons}

Searches by ATLAS and CMS for $H,\ A\to\tau^+\tau^-$ now exclude a large portion of the
$m_A\ vs.\ \tan\beta$ plane~\cite{bib:CMS_SUSYHiggs_tautau,bib:ATLAS_SUSYHiggs_tautau}. 
In particular, the region around $\tan\beta\sim 50$, 
which is favored by Yukawa-unified SUSY GUT theories, now excludes $m_A<500$ GeV.
For $\tan\beta =10$, the range $120\ {\rm GeV}<m_A<220$ GeV is excluded.
ATLAS excludes charged Higgs bosons produced in association with a $t \bar{t}$ pair
for masses below about $150$~GeV for $\tan{\beta} \sim 20$~\cite{bib:ATLAS_SUSYHiggs_taunu}.

\subsubsection{Impact of Higgs searches on SUSY models}

\begin{figure}[htb]
  \begin{center}
\includegraphics[width=0.45\textwidth]{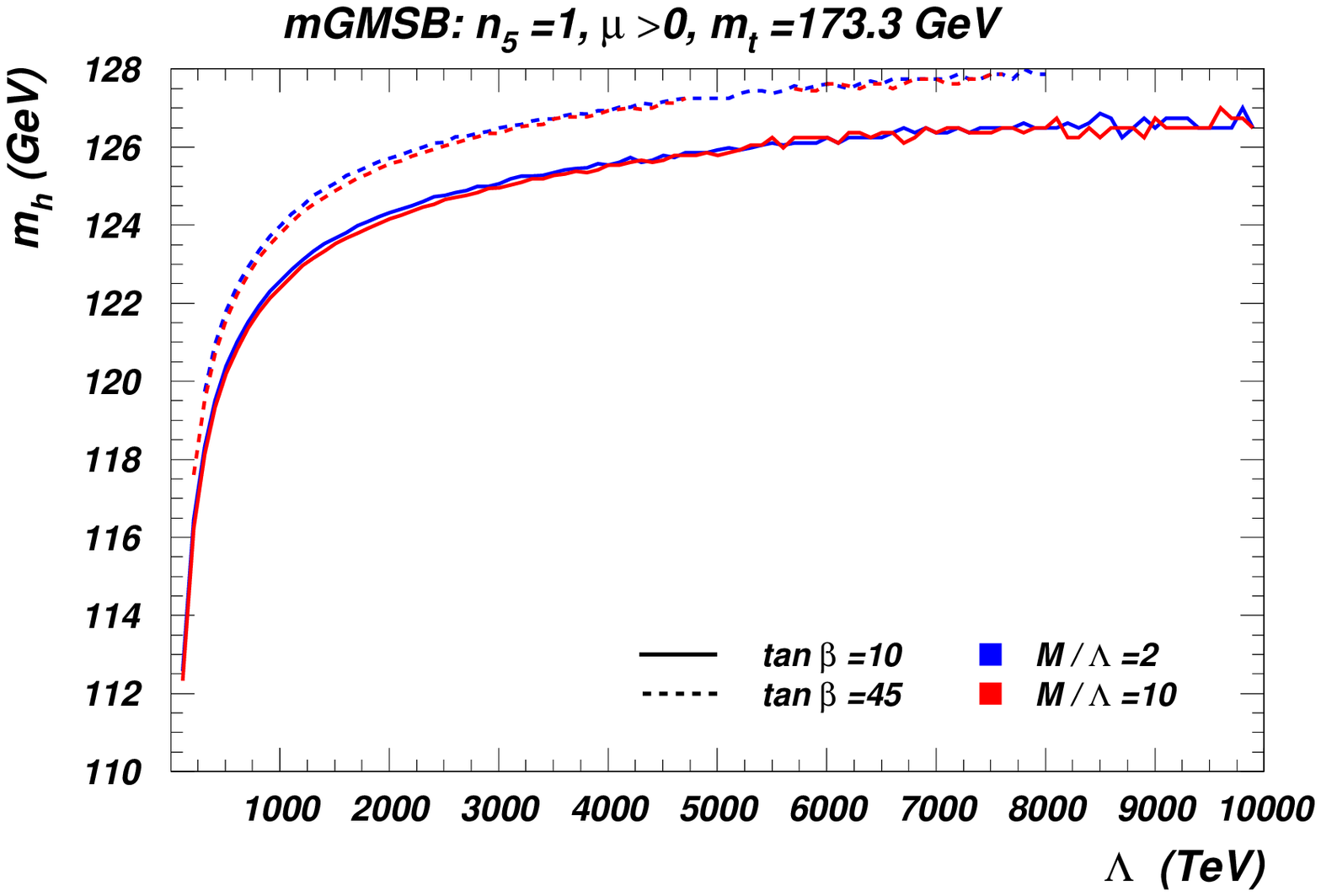}
\hspace{0.1cm}
\includegraphics[width=0.45\textwidth]{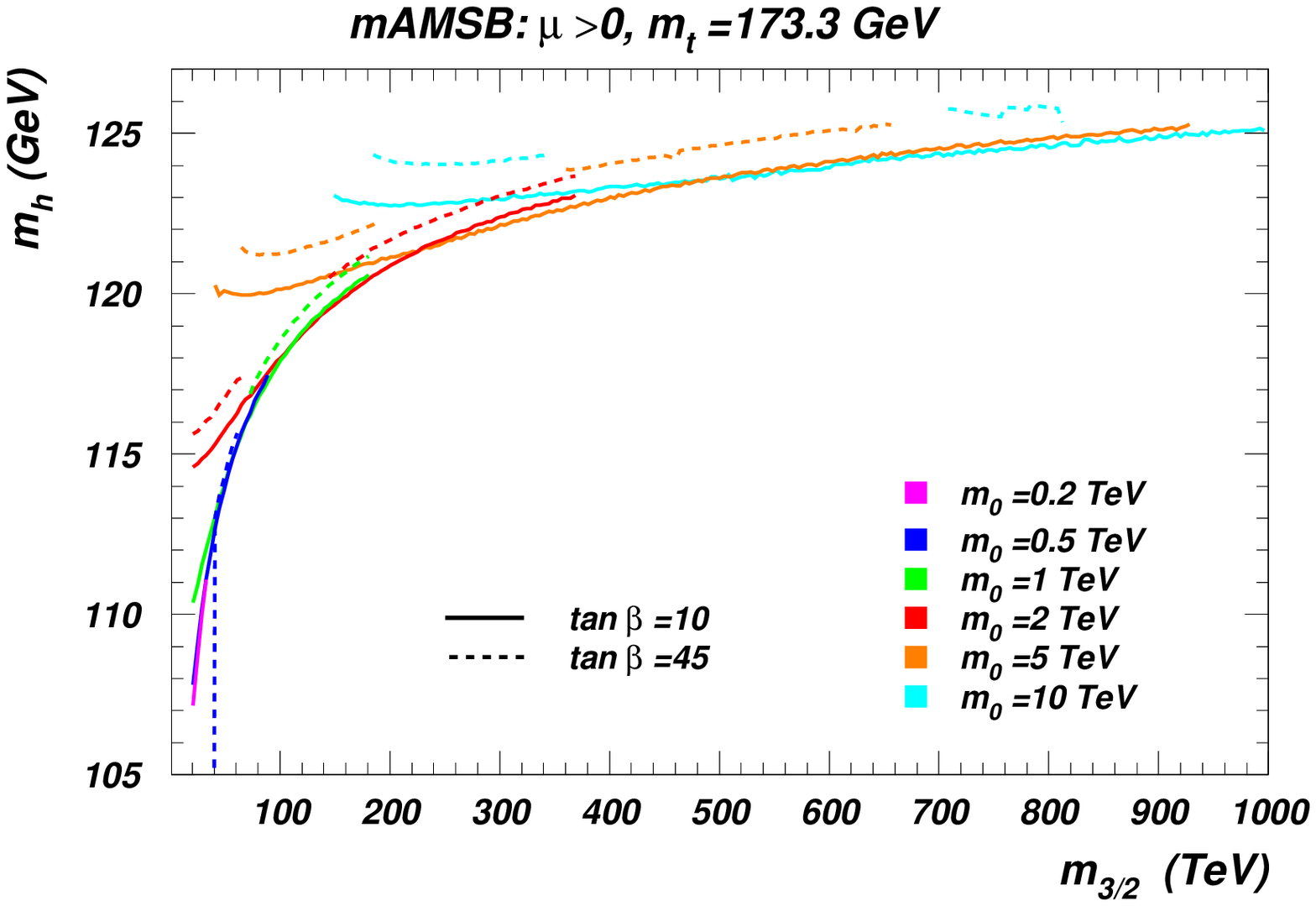}
  \end{center}
  \caption{Value of $m_h$ in mGMSB and in mAMSB versus $\Lambda$ and $m_{3/2}$
from~\cite{Baer:2012uy}.}
\label{fig:Mh_GMSB_AMSB}
\end{figure}

A Higgs mass of $m_h =125\pm 3$~GeV lies below the value of $m_h\sim 135$ GeV which is allowed
by calculations within the MSSM. 
However, such a large value of $m_h$ requires large radiative corrections and 
large mixing in the top squark sector. In models such as mSUGRA, trilinear soft parameters $A_0\sim \pm 2m_0$
are thus preferred, and values of $A_0\sim 0$ would be ruled out~\cite{Baer:2011ab,Heinemeyer:2011aa}.
In other constrained models such as the minimal versions of GMSB or AMSB,
Higgs masses of $125$~GeV require even the lightest of sparticles to be in the multi-TeV range~\cite{Baer:2012uy}, 
as illustrated in Figure~\ref{fig:Mh_GMSB_AMSB}.
%

In the mSUGRA/CMSSM model, requiring a Higgs mass of about $125$~GeV pushes the best fit point in
$m_0$ and $m_{\frac{1}{2}}$ space into the multi-TeV range~\cite{Baer:2011ab} and makes global fits 
of the model to data increasingly difficult~\cite{Buchmueller:2011ab}. This has provided motivation
for extending the MSSM with gauge singlets~\cite{Hall:2011aa,King:2012is} or vector-like matter~\cite{Martin:2009bg} 
both of which allow for somewhat heavier values of $m_h$.
 
\subsection{Review of sparticle searches at LHC}

\subsubsection{Gluinos and first/second generation squarks}

The ATLAS and CMS collaborations have searched for multi-jet$+\eslt$ events 
arising from gluino and squark pair production in 4.4 fb$^{-1}$ of 2011 data taken 
at $\sqrt{s}=7$~TeV~\cite{bib:ATLAS_jets, bib:CMS_razor}. 
In the limit of very heavy squark masses, they exclude $m_{\tg}\alt 0.8$~TeV, while for 
$m_{\tq}\simeq m_{\tg}$ then $m_{\tg}\alt 1.4$~TeV is excluded. Here, $m_{\tq}$ refers to a 
generic first generation squark mass scale, since these are the ones whose production rates 
depend strongly on valence quark PDFs in the proton.

Both collaborations in addition have searched for gluino and squark cascade 
decays~\cite{Baer:1986au} 
assuming more specific decay chains leading to signatures involving leptons and photons as well as $b$-jets~\cite{bib:ATLAS_jetslepton, bib:ATLAS_multijets, bib:ATLAS_GMSB_staus, bib:ATLAS_bjets, bib:ATLAS_SSleptons, bib:ATLAS_2photons, bib:CMS_photons, bib:CMS_bjets_SSleptons, bib:CMS_JZB, bib:CMS_multileptons, bib:CMS_jetslepton, bib:ATLAS_4leptons}.
In most cases, the limits on the gluino mass are rather similar to the ones
from the multi-jet$+\eslt$ analyses, with values of $m_{\tg}\alt 0.8-1$~TeV being excluded depending on the particular decay chain. 

If the gluino decays dominantly into third generation squarks, the gluino mass limits are somewhat
weaker, typically in the range of $0.65$ to $0.8$~TeV, again depending on the exact decay 
chain~\cite{bib:ATLAS_bjets, bib:CMS_bjets_SSleptons, bib:ATLAS_SSleptons}. 
These results are soon expected to be upgraded to include the full 5 fb$^{-1}$ data set.

Some analyses have addressed the situation where there are small mass differences between 
mother and daughter particles in the decay chain. In one case, ATLAS considered gluino decays via 
an intermediate chargino~\cite{bib:ATLAS_jetslepton}. Using a soft-lepton tag, they reach down to 
$\tg -\tz_1$ mass differences of $\sim 100$~GeV. In this case, gluino masses are only excluded up to $0.5$~TeV.

\subsubsection{Sbottom and Stop}

A recent ATLAS search for direct bottom squark pair production 
followed by $\tb_1\to b\tz_1$ decay ($pp\to\tb_1\bar{\tb}_1\to b\bar{b}+\eslt$)
based on 2 fb$^{-1}$ of data at $\sqrt{s}=7$ TeV now excludes 
$m_{\tb_1}\alt 350$ GeV for $m_{\tz_1}$ as high as $120$~GeV. For larger values of 
$m_{\tz_1}$, the limit vanishes at present~\cite{bib:ATLAS_dir_sbottom}.
These limits also apply to top squark pair production where $\tst_1\to b\twp$ decay and
the $\twp$ decays to soft, nearly invisible particles, as would be expected in natural SUSY.
From a search for events with $b$-jets, same-sign di-leptons and missing energy, 
CMS puts a limit on the mass of directly produced bottom squarks to be larger than $370$~GeV for chargino masses 
between $100$ and $200$~GeV and a $\tz_1$ mass of $m_{\tz_1} = 50$~GeV~\cite{bib:CMS_bjets_SSleptons}.

In the context of GMSB with the $\tz_1$ as higgsino-like NLSP and a gravitino $\tG$ LSP, 
ATLAS searched for direct top squark pair production, followed by $\tst_1\to b\twp$ or, 
when kinematically allowed, also $t\tz_1$. 
Based on $2$~fb$^{-1}$, they exclude top squark masses up to $330$~GeV for NLSP masses around $190$~GeV~\cite{bib:ATLAS_GMSBstop}. 
This limit relies on the GMSB specific decay of the $\tz_1$ into $Z \tG$, especially on two (same flavour, opposite sign) leptons 
consistent with the $Z$ mass.

\subsubsection{Electroweakinos}

In models with gaugino mass unification and heavy squarks (such as mSUGRA with large $m_0$), 
electroweak gaugino pair production $pp\to\twpm_1\tz_2$ is the dominant SUSY particle production cross section at 
LHC7 for $m_{\tg}>0.5$ TeV\cite{Baer:2012wg}. 
If the $\twpm_1$ and $\tz_2$ decay leptonically and $\tz_2\to\tz_1 Z$ decay is closed, then this reaction leads to the well-known 
trilepton plus $\eslt$  final state~\cite{Baer:1985at,Baer:1994nr} which may be observable over SM backgrounds. 
A search by ATLAS using 2.1 fb$^{-1}$ of data~\cite{bib:ATLAS_ewkino} 
has been interpreted in the pMSSM and in a simplified model assuming chargino and neutralino decay to 
intermediate sleptons, which enhances the leptonic branching fractions. 
In the simplified model case, $m_{\twpm_1}<250-300$ GeV are ruled out for $m_{\tz_1}=0-150$ GeV. 
In the pMSSM as well as in the simplified model interpretation it is assumed that the lighter set of sleptons, 
including the third generation, is mass degenerate and fulfils $m_{\tl}=(m_{\twpm_1}-m_{\tz_1})/2$, 
which maximizes the lepton momenta and thus the acceptance.
Thus this analysis does in particular not apply to scenarios with a small $\ttau_1$-$\tz_1$ mass difference, 
which are still a viable scenario even for $M_2$ and $\mu$ values depicted as excluded in~Fig.~2 of reference~\cite{bib:ATLAS_ewkino}. 
Furthermore, the theoretically more interesting case of chargino and neutralino three-body leptonic decay through $W^*$ and $Z^*$ 
should be possible with 10-20 fb$^{-1}$ of data, as should  the trilepton signal from $pp\to\twpm_1\tz_2\to WZ+\eslt$~\cite{Baer:2012wg}.

\subsubsection{Electroweakinos with extremely small mass differences}

In models such as AMSB where the light chargino $\twpm_1$ and neutralino $\tz_1$ are expected to be wino-like,
the expected $\twpm_1-\tz_1$ mass gap is expected to be $\sim 100-200$ MeV. Such a small mass gap
implies the $\twpm_1$ will actually fly a short but possibly observable distance before decaying into
very soft pion(s). A search by ATLAS using 4.7 fb$^{-1}$ has been made for long lived charginos arising from
gluino and squark cascade decays~\cite{bib:ATLAS_disapptrk}. 
Thus, the search looks for three high $p_T$ jets plus $\eslt >130$ GeV.
Within this event class, a search is made for events with hits in the transition radiation tracker (TRT)
which ultimately disappear. No signal is seen above expected background levels, leading to 
limits on $m_{3/2}>32$ TeV in the mAMSB model. More generally, lifetimes between
$\tau_{\twpm_1}\sim 0.2-90$ ns are excluded for $m_{\twpm_1}<90$ GeV at 95\% CL.  

\subsubsection{Heavy stable charged particles}

Long-lived quasi-stable charged or colored particles are common in many versions of
supersymmetric models. Examples include GMSB models with a $\ttau_1$ as NLSP which decays
to $\tau +\tG$, or models such as split SUSY where gluino decays are suppressed by an ultra-heavy
squark mass scale. In the latter case, any quasi-stable gluinos  which are produced at LHC
would be expected to hadronize into a gluino hadron, which could be either charged or neutral.

A search by ATLAS using 2.1 fb$^{-1}$ of data looks for anomalous $dE/dx$ energy loss measurements
in the Pixel detector. Since no deviation from expected background levels was found, 
they were able to exclude the production of gluino hadrons with $m_{\tg}<810$ GeV~\cite{bib:ATLAS_hscp_pixel}.

\subsubsection{$R$-Parity Violation}

The ATLAS collaboration has searched for $R$-parity violating SUSY (for a review, see~\cite{Dreiner:1997uz}) in the context of the
mSUGRA/CMSSM model in two scenarios. 

In the case that $m_0\sim 0$, the tau-slepton $\ttau_1$ is the LSP. To be compatible with cosmological bounds on relic stable charged
 particles produced in the Big Bang, it is assumed that $\ttau_1$ decays to $\tau e^\mp (\ell^\pm\nu_\ell )$ where $\ell =e$ or 
 $\mu$ via the $R$-parity coupling $\lambda_{121}$. A search for four isolated leptons plus $\eslt$
in $2$~fb$^{-1}$ of data allows them to exclude $m_{1/2}<800$ GeV at 95\% CL for $\tan\beta <40$ and 
$m_{\ttau_1}>80$ GeV~\cite{bib:ATLAS_RPV_stau}. 

Furthermore, ATLAS has published an interpretation of their search for events with one lepton, jets and missing transverse 
energy in $1$~fb${^{-1}}$ of data~\cite{ATLAS:2011ad} in the context of bilinear $R$-parity violating SUSY, where the bRPV 
parameters are determined by fitting them to neutrino oscillation data~\cite{Porod:2000hv}. For $\tan{\beta}=10$, $A_0 = 0$ 
and $\mu>0$, they exclude values of $m_0$ up to $430$~GeV for $m_{1/2} = 290$~GeV. For smaller or larger values of $m_{1/2}$ 
the exclusion in $m_0$ is weaker; values of $m_{1/2}<240$~GeV have not been studied at all.

\section{Implications for ILC and benchmark points}
\label{sec:BMs}

The results from the previous sections, when summarized, yield the following grand picture:

\begin{itemize}
\item{\bf Squarks and gluinos:} Ironically, the strongest LHC limits on sparticle masses apply to 
the first generation squarks and gluinos, while these are the most remotely connected to the determination of 
the electroweak scale, and to the weak boson masses. So while $m_{\tg}\agt 1.4$ TeV 
for $m_{\tq}\sim m_{\tg}$, these limits hardly affect naturalness limits, which prefer $m_{\tg}\alt 3-4$ TeV
and basically do not constrain first generation squarks, so that $m_{\tq}$ values into the tens of
TeV regime are certainly allowed. 
\item {\bf Electroweakinos:} The masses of the electroweakinos -- constrained by LEP2 to have
$m_{\tw_1}>103.5$ GeV -- are hardly constrained by LHC7 data unless they are connected with 
1. a light gluino (via the gaugino mass unification assumption) or first/second generation squarks 
allowing for strong production or 2. in conjuction with light sleptons appearing in the 
electroweakino decay right in between the $\tz_1$ and $\tz_2, \twpm_1$ masses. 
In particular, $m_{\tz_1}$, $m_{\tz_2}$ and $m_{\twpm_1}$ can very well be below $200$~GeV 
as motivated by naturalness. 
Very likely they have at least a sizable Higgsino component, and thus could very well have small mass 
splittings. Several of the scenarios proposed below exhibit such a pattern for the light electroweakinos. 
The heavier electroweakinos are likely not directly observable at the ILC. 
The proposed benchmarks cover various options in this respect. 
\item {\bf Sleptons:} The most important indication for light sleptons is still $(g-2)_\mu$. 
They are so far not constrained directly by LHC7 data (but see~\cite{Baer:1993ew} for projections). 
If a common matter scalar mass $m_0$ at the GUT scale is assumed, 
then the stringent LHC7 bounds on first and second generation squarks imply also rather 
heavy sleptons. 
Most of the scenarios below have heavy sleptons and thus do not explain the $(g-2)_\mu$ anomaly. 
If non-universality of matter scalars is assumed, then the slepton masses are completely unconstrained 
and all sleptons could still lie within reach of the ILC, as illustrated by the $\delta M\ttau$ and NMH benchmarks described 
below: both these scenarios allow for perfect matches to the observed  $(g-2)_\mu$ value. 
In natural SUSY -- while the first two slepton generations are expected to be heavy -- 
the $\ttau_1$ can be quite light due to the limited mass of the top squarks. 
\item {\bf Third generation squarks:} Direct limits on the third generation squarks from LHC7 
are far below those for the first generation, so that especially the top squark could very well be in
the regime expected from naturalness and thus accessible at the ILC. 
Both the natural SUSY benchmark and the $\delta M\ttau$ benchmark described in Subsections~\ref{sec:NS} 
and~\ref{sec:tdr1} give examples with light $\tst_1$ and possibly $\tb_1$ and $\tst_2$.
\item {\bf SUSY Higgses:} The possibly SM-like properties of a $125$~GeV Higgs scalar, as hinted at 
by LHC7 data, suggests that the other SUSY Higgses could be rather heavy, although of course 
a firm statement in this regard will require not only a Higgs discovery but also precise measurements of the branching ratios. 
We present in section~\ref{sec:nuhm2} a NUHM2 scenario with light $A$, $H$ and $H^\pm$; also,
the $\delta M\ttau$ benchmark features heavy Higgses which should be observable at a $1$~TeV 
$e^+e^-$ collider. 
\end{itemize} 

Based on these observations, we propose a set of benchmark points which can be used to 
illustrate the capabilities of ILC with respect to supersymmetry, 
and for future optimization of both machine and detector design. 
The suggested points all lie outside the limits imposed by LHC7 searches. 
Some of these scenarios might be discoverable or excluded by upcoming LHC8 searches, while others
will be extremely difficult to detect at LHC even with $3$~ab$^{-1}$ of data at $\sqrt{s}=14$~TeV.
The spectra for all benchmarks are available online~\cite{bib:webpage} in the SUSY Les Houches Accord format.

\subsection{Natural SUSY}
\label{sec:NS}

Natural SUSY (NS) models are characterized by~\cite{Brust:2011tb,Papucci:2011wy,Baer:2012uy2}: 
\bi
\item a superpotential higgsino mass parameter $\mu <\Lambda_{NS}\sim 200$ GeV, 
\item a sub-TeV spectrum of third generation squarks $\tst_1$, $\tst_2$ and $\tb_1$, 
\item an intermediate scale gluino $m_{\tg}\alt 3-4$ TeV with $m_A\alt |\mu|\tan\beta$ and 
\item multi-TeV first/second generation matter scalars $m_{\tq ,\tell }\sim 10-50$ TeV.
\ei
The last point offers at least a partial decoupling solution to the SUSY flavor and $CP$ problems.  

The suggested model parameter space is given by~\cite{Baer:2012uy2}:
\begin{equation}
m_0(1,2),\ m_0(3),\ m_{1/2},\ A_0,\ \tan\beta,\ \mu,\ m_A \;.
\label{eq:NSparams}
\end{equation}
Here, we adopt a NS benchmark point as calculated using Isasugra 7.82~\cite{Paige:2003mg} with parameters
$m_0(1,2)=13.5$ TeV, $m_0(3)=0.76$ TeV, $m_{1/2}=1.38$ TeV, $A_0=-0.167$ TeV, 
$\tan\beta =23$ GeV, $\mu =0.15$ TeV and $m_A=1.55$ TeV.
The resulting mass spectrum is listed in Table~\ref{tab:bm1}.

Due to their small mass differences, the higgsino-like light electroweakinos will tend to look 
like missing transverse energy to the LHC. 
The next heavier particle is the $\tst_1$. 
Since the mass difference $m_{\tst_1}-m_{\tz_1}$ is less than the top mass, the decay 
$\tst_1\to b \twpm_1$ dominates, thus making the signature for $\tst_1$ pair production 
two acollinear $b$-jets plus missing transverse energy.

For ILC, the spectrum of higgsino-like $\twpm_1$, $\tz_1$ and $\tz_2$ will be accessible 
for $\sqrt{s}\agt 320$ GeV via $\twpm_1$ and $\tz_2\tz_2$ pair production and $\tz_1 \tz_2$ mixed production, 
albeit with a mass gap $m_{\twpm_1}-m_{\tz_1}\sim m_{\tz_2}-m_{\tz_1}\sim 7.5$ GeV: thus, 
visible energy released from decays will be small.
Specialized cuts allowing for ILC detection of light higgsinos with small mass gaps 
have been advocated in Ref's~\cite{Baer:2003ru} and \cite{Baer:2004zk}; there it is also demonstrated
that ILC will be able to measure the values of $\mu$ and $M_2$ and show that $|\mu |<M_2$. 

In the case of very small mass gaps, a hard ISR photon radiated from the initial state may help to lift 
the signal out of the substantial background of photon--photon induced processes. The experimental performance 
of this ISR recoil method has been evaluated recently in full simulation of the ILD detector in context of radiative
WIMP / neutralino production~\cite{Bartels:2012ui, Bartels:2012rg}.
The cross-sections are typically in the few tens of fb region~\cite{Baer:2011ec} 
and thus should be detectable in the clean ILC environment. 
Similar signatures have also been investigated in the context of AMSB for the 
TESLA TDR~\cite{AguilarSaavedra:2001rg}.

As $\sqrt{s}$ is increased past $600-800$ GeV, then also $\tst_1\bar{\tst}_1$, $\tnu_\tau\bar{\tnu}_\tau$ 
and $\ttau_1\bar{\ttau}_1$ become successively accessible. 
This benchmark model can be converted to a model line by
varying the GUT-scale third generation mass parameter $m_0(3)$ or by varying $\mu$.
The light higgs mass $m_h$ can be pushed as high as $\sim 124$ GeV if larger values
of $m_0(3)$ and $|A_0|$ are selected~\cite{Baer:2012uy2}.  

\subsection{Hidden SUSY}
\label{sec:hs}

Models of ``hidden SUSY''~\cite{Baer:2011ec} are motivated by the fact that the 
magnitude of the superpotential higgsino mass parameter $\mu$ itself
has been suggested as a measure of fine-tuning~\cite{Chan:1997bi}. This idea has been used to
argue that mSUGRA/CMSSM models in the hyperbolic branch/focus point region
are less fine-tuned than generic parameter space regions. 
Natural SUSY models wherein $\Lambda_{NS}$ slides into the $\agt 300$ GeV regime
but maintain $|\mu |\alt 300$ GeV morph into hidden SUSY.
Here, we select a model where all mass parameters are large except for the $\mu$ 
parameter, which may lie in the $100-300$ GeV range. 
The parameter space suggested is that of the 2-parameter non-universal
Higgs mass (NUHM2) model:
\begin{equation}
m_0,\ m_{1/2},\ A_0,\ \tan\beta,\ \mu,\ m_A \;.
\label{eq:nuhm2}
\end{equation}
Here, we adopt a benchmark point with parameters
$m_0=5$ TeV, $m_{1/2}=0.8$ TeV, $A_0=-8.3$ TeV, $\tan\beta =10$
with $\mu =0.15$ TeV and $m_A=1$ TeV. The spectrum is given in Table~\ref{tab:bm1}.

Hidden SUSY models are very difficult to detect at LHC. 
In contrast to natural SUSY, the third generation scalars are also beyond $1$~TeV. 
While the higgsino-like light charginos and neutralinos are produced at large rates, 
the very low energy release from their decays will be hard to detect above background levels, 
making them all look like missing transverse energy. 
If the cross-sections are large enough, the decays of the $\tz_3\to\twpm_1 W^{\pm}$, $\tz_{1,2} Z$ 
or $\tz_{1,2} h$ might provide a source of isolated leptons visible at the LHC 
if the $\tst_1$ is too heavy for detection.

The ILC operating at energy $\sqrt{s}\agt 300$ GeV should be able to detect and distinguish
$\twp_1\twm_1$ and $\tz_1\tz_2$ production as in the natural SUSY case discussed above. 
The small mass gap, angular distribution and polarization dependence of the signal cross sections 
may all be used to help establish the higgsino-like nature of the light $\twpm_1$, $\tz_2$ and $\tz_1$. 
In addition, the $\tz_3$ is accessible in mixed production with the lighter neutralinos already at 
$\sqrt{s}\agt 500$~GeV.

Phenomenologically similar scenarios -- which are even more minimal case in the sense that the $\tz_3$ 
and the $\tst_1$ are in the multi-TeV regime as well -- 
have been suggested by Br\"ummer and Buchm\"uller~\cite{Brummer:2012zc}. 
We will discuss one example in section~\ref{sec:bb}. 

\subsection{NUHM2 benchmark with light $A$, $H$ and $H^\pm$}
\label{sec:nuhm2}

This benchmark point, constructed within the 2-parameter
non-universal Higgs model (NUHM2), provides a model with
relatively light $A$, $H$  and $H^\pm$ Higgs bosons while the remaining sparticles
are beyond current LHC reach. We adopt parameters
$m_0=10$ TeV, $m_{1/2}=0.4$ TeV, $A_0=-16$ TeV, $\tan\beta = 6$
with $\mu =5$ TeV and $m_A=275$ GeV. The values of $m_h=124.4$ GeV, 
with $m_H=277.5$ GeV and $m_{H^+}=286.0$ GeV are obtained with FeynHiggs~\cite{bib:feynhiggs}. 
The only colored sparticles accessible to the LHC are the gluinos with $m_{\tg}=1.225$ TeV, 
while most squarks live at around $m_{\tq}\sim 10$ TeV. 
The gluino decays are dominated by $\tg \ra \tz_1 t \bar{t}$ and  $\tg \ra (\twpm _1 \ra \tz_1 W^{\pm}) t b$,
and thus will require dedicated analyses for high multiplicity final states or boosted 
techniques for identifying $W$- or $t$-jets. 
The signal $pp\to\tw_1\tz_2\to Wh+\eslt\to \ell\nu_\ell+b\bar{b}+\eslt$ should ultimately be
observable at LHC14~\cite{Baer:2012ts}.
The Higgs bosons, apart from the light $CP$-even one, can most probably not be observed at the LHC 
in this low $\tan{\beta}$ and $m_A$ region~\cite{ATLASblueplot}.

At the ILC with $\sqrt{s}\sim 0.5$ TeV, we expect $e^+e^-\to Ah,\ ZH$ to occur at observable rates. 
As $\sqrt{s}$ rises beyond $600$~GeV, $AH$ and $H^+H^-$ production becomes accessible
while mixed $\tz_1 \tz_2$  pair production, though accessible, is suppressed. 
At $800$~GeV,  $\twpm_1$ pairs will be produced in addition. 
Due to heavy sleptons and the sizable mass gap between $\twpm_1,\ \tz_2$ and the $\tz_1$, 
one expects electroweakino decays to real $W^{\pm}$ and $Z$ bosons, 
very similar to the ``Point~5'' benchmark studied in the Letter of Intents 
of the ILC experiments~\cite{Abe:2010aa, Aihara:2009ad}.

\subsection{mSUGRA/CMSSM}

Large portions of mSUGRA model parameter space are now ruled out by
direct searches for gluino and squark production at LHC7.
In addition, if one requires $m_h\sim 124-126$ GeV, then even larger portions of
parameter space are excluded: $m_{1/2} < 1$~TeV (corresponding to $m_{\tg}<2.2$ TeV)
for low $m_0$ and $m_0<2.5$ TeV (corresponding to $m_{\tq}<2.5$ TeV) for low $m_{1/2}$~\cite{Baer:2011ab}. 
These tight constraints rule out almost all of the co-annihilation and 
$A$-funnel annihilation regions~\cite{Baer:2011ab,Baer:2012uy}.
The HB/FP region moves to very large $m_0>10$ TeV since now $|A_0|$ must be large to accommodate 
the rather large value of $m_h$. Some remaining dark matter allowed parameter space thus remains.

An example is provided by an mSUGRA benchmark point with 
$m_0=15.325$ TeV, $m_{1/2}= 0.845884$ TeV, $A_0=-10.8126$ TeV and $\tan\beta =20.197$. 
The masses are shown in Table \ref{tab:bm1}.
At this point, $m_{\tg}=2320$ GeV and $m_{\tq}\sim 15.3$ TeV. However, $\mu\sim 145$ GeV,
and so $m_{\twpm_1}=155.3$ GeV and $m_{\tz_2}=154.8$ GeV and $m_{\tz_1}=141.6$ GeV. Thus, this point --
although very fine-tuned in the EW sector (with $m_{\tst_1}\sim 8.7$ TeV) -- would allow
$\twp_1\twm_1$ and $\tz_1\tz_2$ production at ILC with a $\twpm_1-\tz_1$ mass gap of 14 GeV. The $\tz_1$ would be of mixed bino-higgsino variety. When increasing $\sqrt{s}$ towards $1$~TeV, the heavier neutralinos become accessible in mixed production and $\tz_3$ pair production.

Since all scalars are above $10$~TeV (apart from the lighter top squark at $m_{\tst_1} \sim 8$~TeV), the most promising 
signature for the LHC is gluino production, followed 
by $\tg \ra \tz_i t \bar{t}$ and  $\tg \ra (\twpm _j \ra \tz_1 W^{\pm}) t b$ as discussed in case of the NUHM2 benchmark in
Section~\ref{sec:nuhm2}.

\subsection{Model with non-universal gaugino masses (NUGM)}

In supergravity, gaugino masses arise from the Lagrangian term (using 4-component spinor notation)
\be
{\cal L}_F^G = -{1\over 4}e^{G/2}\frac{\partial f_{AB}^*}{\partial\hat{h}^{*j}}\left|_{\hat h\to h}\right.
\left( G^{-1}\right)_k^jG^k\bar{\lambda}_A\lambda_B
\ee
where $f_{AB}$ is the holomorphic gauge kinetic function with gauge indices $A,\ B$ 
in the adjoint representation, $\lambda_A$ are four-component gaugino fields and the $\hat{h}_m$ are hidden sector fields 
needed for breaking of supergravity. 
If $f_{AB}\sim \delta_{AB}$, then gaugino masses are expected to be universal at the high energy scale where SUSY breaking takes place. 
However, in general supergravity, $f_{AB}$ need only transform as the symmetric product of two adjoints. 
In general, gaugino masses need not be universal at any energy scale, giving rise to models with non-universal gaugino masses (NUGM).

For a NUGM benchmark, we select a model with $m_0=3$ TeV, $A_0=-6$ TeV, $\tan\beta =25$ 
and $\mu >0$. We select gaugino masses at the GUT scale as $M_1=M_2=0.25$ TeV with $M_3=0.75$ TeV. 
The spectrum is listed in column 6 of Table~\ref{tab:bm1}. 
With $m_{\tg}\simeq 1.8$ TeV and $m_{\tq}\simeq 3$ TeV, the model is clearly beyond current LHC reach for gluinos and squarks.
The model should be testable in future LHC searches, not only in with the standard jets plus missing $E_t$ analyses, 
but also via searches tailored for very high multiplicity final states and using $b$-jet tagging~\cite{Kadala:2008uy}, 
since the gluino almost exclusively decays via $\tg \ra \tst_1 t$ followed by $\tst_1 \ra \tz_1 t$. 
In addition, the production channel $pp\to\twpm_1\tz_2\to WZ+\eslt$ may be testable in the near future~\cite{Baer:2012wg}. 

The rather light spectrum of electroweak gauginos with $m_{\twpm_1}\sim 2 m_{\tz_1}\sim 216$ GeV allows for chargino pair production at ILC 
followed by $\twpm_1\to\tz_1 W$ decay, yielding a $W^+W^- +\esl$ signature.
The $\tz_1\tz_2$ and $\tz_2\tz_2$ production channels tend to be suppressed, but may
offer additional search avenues albeit at low rates.

\begin{table}[h!]
\centering
\begin{tabular}{lccccc}
\hline
\hline
{\rm PMQ}                 & NS          & HS   & NUHM2 & mSUGRA & NUGM  \\
\hline
$m_{0}(1,2)$, $m_{0}(3)$  & 13.35, 0.76 & 5.0  & 10.0  & 15.325  & 3.0 \\
$m_{1/2}$ / $M_1,M_2,M_3$ & 1.38        & 0.7  & 0.4   &  0.8459 & 0.25,0.25,0.75 \\
$\tan\beta$               &  23         & 10   & 6    &  20.2   & 25 \\
$A_{0}$                   & -0.167      & -8.3 & -16.0 &  -10.81 & -6.0  \\
\hline
$m_h$                     & 0.121  & 0.125  & 0.124 & 0.126  & 0.125 \\
$m_{A}$                   & 1.55   & 1.0    & 0.275 &  14.22 & 3.268 \\
$m_{H}$                   & 1.560  & 1.006  & 0.277 &  14.31 & 3.289 \\
$m_{H^{\pm}}$             & 1.563  & 1.011  & 0.286 &  14.31 & 3.293 \\
$\mu$                     & 0.15   & 0.15   & 6.0   &  0.144 & 2.36 \\
\hline
$m_{\tilde{g}}$                & 3.27          & 1.79         & 1.225         & 2.32         & 1.835 \\
$m_{\tilde{\chi}^{\pm}_{1,2}}$ &0.156, 1.18    & 0.154, 0.611 & 0.386, 4.9   & 0.155, 0.756 & 0.216, 2.37   \\
$m_{\tilde{\chi}^0_{1,2}}$     & 0.148, 0.156  & 0.14, 0.158  & 0.192, 0.384 & 0.141, 0.155 & 0.109, 0.215  \\
$m_{\tilde{\chi}^0_{3,4}}$    &0.615, 1.18     &0.32, 0.621   & 4.93, 4.93   & 0.397, 0.780 & 2.36, 2.36 \\
\hline 
$m_{ \tilde{u}_{L,R}}$  &13.58, 13.59          & 5.12, 5.27  & 9.92, 10.21   & 15.31, 15.36 & 3.30, 3.31 \\
$m_{\tilde{t}_{1,2}}$   &0.286, 0.914          & 1.21, 3.55  & 4.14, 7.43    & 8.75, 12.29 & 1.11, 2.29 \\
\hline 
$m_{ \tilde{d}_{L,R}}$       &13.6, 13.6     & 5.12, 5.09   & 9.92, 9.89 & 15.31, 15.37 & 3.30, 3.31  \\
$m_{\tilde{b}_{1,2}}$        &0.795, 1.26    & 3.58, 5.0    & 7.45, 9.84  & 12.26, 14.85 & 2.30, 2.99   \\
\hline
$m_{ \tilde{e}_{L,R}}$         &13.4, 13.3  & 5.11, 4.8  & 10.2, 9.66 & 15.31, 15.31 & 3.0, 3.0   \\
$m_{\tilde{\tau}_{1,2}}$       &0.43, 0.532 & 4.73, 5.07 & 9.61, 10.1 & 14.68, 14.99 & 2.6, 2.81  \\
\hline
$\Omega_{\tz}^{std}h^2$        &0.007       & 0.009     & 210         & 0.008 & 1540 \\
$\langle\sigma v\rangle(v\rightarrow 0)\ [\mathrm{cm^3/s}]$ & 3.1$\times 10^{-25}$ & 2.8$\times 10^{-25}$ & 5.1$\times 10^{-30}$  & 2.9$\times 10^{-25}$ & 1.5$\times 10^{-32}$  \\
$\sigma^{SI}(\tz p)\times 10^{9}$ [pb]  & 2.0 & 11. & 0.007 & 4.0 & 0.0004 \\
\hline
$a_\mu^{SUSY} \times 10^{10}$  & 0.03   & 0.09 & 0.05 & 0.02 & 0.45 \\
$BF(b\rightarrow s\gamma )\times 10^4$  & 3.3 & 3.3 & 3.48 & 3.05 & 2.95 \\
$BF(B_S\rightarrow \mu^+\mu^- )\times 10^9$  & 4.2 & 3.8 & 3.9 & 3.8 & 3.9  \\
$BF(B_u\rightarrow \tau\nu_\tau )\times 10^4$  & 1.3 & 1.3 & 1.3 & 1.3 & 1.3  \\
\hline
\hline
\end{tabular}
\caption{Input parameters and mass spectrum and rates for post LHC7
benchmark points $1-5$. All masses and dimensionful parameters are in TeV units. 
All values have been obtained with Isasugra apart from Higgs masses for the NUHM2 point, 
which have been taken from FeynHiggs.
}
\label{tab:bm1}
\end{table}

\subsection{A pMSSM model with light sleptons}
\label{sec:tdr1}

In many constrained SUSY models where slepton and squark masses are
correlated at some high energy scale, relatively light sleptons with 
mass $\sim 100-200$ GeV are forbidden. However, if we invoke the greater 
parameter freedom of the pMSSM, then spectra with light sleptons and heavy squarks
can easily be generated. In fact, these models have some degree of motivation in that
they naturally reconcile the measured $(g-2)_\mu$ anomaly 
(which favors light smuons) with the measured $b\to s\gamma$ branching fraction 
(which favors rather heavy third generation squarks).

In the pMSSM\cite{Baer:1993ae,Djouadi:2002ze}, one inputs {\it weak scale} values of the following parameters:
1. $m_{\tg},\mu ,m_A,\tan\beta$, 2. $m_Q,m_U,m_D,m_L,m_E$ for each of the three generations,
3. gaugino masses $M_1$ and $M_2$ and 4. third generation trilinear $A_t,A_b$ and $A_\tau$.
This gives a 19 dimensional parameter space if first and second generation scalar masses are taken as
degenerate, else a 24 dimensional parameter space for independent first, second and 
third generations.\footnote{ 
Alternatively, the $SU(3)$ gaugino mass $M_3$ may be substituted for the physical
gluino mass as an input.}
As an example, we specify the ``$\delta M\ttau$'' benchmark with the following parameters, 
all given at a scale of $1$~TeV:

\begin{itemize}
\item Higgs sector parameters:\\
 $\tan (\beta) = 10$, $\mu = 200$~GeV, $m_A = 400$~GeV,
\item trilinear couplings: $A_t=A_b=A_\tau = -1.8$~TeV,
\item gaugino mass parameters:\\
 $M_3 = 2$~TeV, $M_2 = 225$~GeV, $M_1 = 107$~GeV,
\item slepton mass parameters:\\
 $m_L(1,2,3) = 200$~GeV, $m_E(1,2) = 125$~GeV, 
$m_E(3) = 103$~GeV,
\item squark mass parameters:\\
 $m_Q(1,2) = m_D(1,2) = m_U (1,2) = 2$~TeV, 
 $m_L(3) = 1.5$~TeV,  $m_U(3) = m_D(3) =  400$~GeV.
\end{itemize}

The resulting sparticle masses, which have been obtained with SPheno~\cite{Porod:2003um, Porod:2011nf} 
with Higgs masses calculated by FeynHiggs~\cite{bib:feynhiggs}, along with the neutralino relic density obtained
from~\cite{bib:micromegas}, are listed in Table~\ref{tab:bm2}.

With masses around  $2$~TeV, the gluino and the partners of the light quarks are beyond current LHC limits, 
especially since the gluino decays dominantly via $\tst_1 t$ or $\tb_1 b$. 
Although light sleptons are present, the current limits on direct electroweakino 
production~\cite{bib:ATLAS_ewkino} do not cover this case due to the small mass difference between the 
$\ttau_1$ and the $\tz_1$, which leads to soft $\tau$ leptons in the chargino and 
neutralino decays instead of the searched for high $p_t$ electrons and muons.

All sleptons and electroweakinos are within ILC reach at $\sqrt{s} \alt 500$~GeV. 
In addition, the light top and bottom squarks as well as the heavy Higgs bosons
would be accessible at ILC with $\sqrt{s}\sim 1$~TeV. 


Due to the large number of production processes open already at $\sqrt{s}\sim 500$~GeV, 
which often yield long cascades\cite{Baer:1988kx}, $\delta M\ttau$ is actually an experimentally challenging scenario for ILC. 
Therefore, it is an ideal case study to demonstrate the separation of many closely spaced new matter states 
with all the tools offered by ILC, including threshold scans and different beam 
polarization configurations, but also taking into account realistic 
assumptions on the beam energy spectrum, accelerator backgrounds and detector resolutions. 

At a center-of-mass energy of 1~TeV or above, the rather small mass difference of  
$40$~GeV between the light stop and sbottom as well as the separation of the heavy Higgs 
states will provide additional experimental challenges.

\subsection{Kallosh-Linde or G2MSSM benchmark}

While minimal anomaly-mediation seems on shaky ground due to its prediction of a
light Higgs scalar $m_h\alt 120$ GeV, other similar models have emerged as perhaps more
compelling. One of these models -- by Kallosh and Linde 
(the KL model~\cite{Kallosh:2004yh,Linde:2011ja}) -- attempts to stabilize
stringy moduli fields via a generalization of the KKLT method~\cite{Kachru:2003aw} utilizing a racetrack
superpotential. The moduli in this theory end up superheavy and allow for 
the chaotic inflationary scenario to emerge in supergravity models. In this class of models, 
the various scalar fields have a mass of the order of the gravitino mass, with  $m_{3/2}\sim 100$ TeV. 
The gauginos, however, remain below the TeV scale, and adopt the usual AMSB form. 
Another stringy model by Acharya {\it et al.}~\cite{Acharya:2008zi} known as G2MSSM also predicts multi-TeV scalars. 
In the G2MSSM, the gauginos are again light, typically with $M_2\ll M_1\sim M_3$ so that again a model 
with light wino-like $\tw_1$ and $\tz_1$ emerges.

To model these cases, we adopt the NUHM2 model, but with non-universal gaugino masses, 
with parameters chosen as $m_0=25$ TeV, $m_{1/2}=200$ GeV, $A_0=0$, $\tan\beta =10$ with
$\mu =m_A=2$ TeV. We then set GUT scale gaugino masses to the AMSB form given by
$M_1=1320$ GeV, $M_2=200$ GeV and $M_3=-600$ GeV. The wino-like $\tz_1$ state is
the lightest MSSM particle with mass $m_{\tz_1}=200.07$ GeV while the wino-like lightest 
chargino has mass $m_{\twpm_1}=200.4$ GeV. 
We also have a bino-like $\tz_2$ with $m_{\tz_2}=616.5$ GeV and a gluino with $m_{\tg}=1788$ GeV.
All matter scalars have mass near the 25 TeV scale, and so decouple. The light Higgs
scalar has mass $m_h=125$ GeV.

In this case, gluino pair production may barely be accessible to LHC14 with of order
$10^2$ fb$^{-1}$ of data~\cite{Baer:2003wx}. At ILC, the decay products from chargino decay will
be extremely soft. However, the wino-like chargino is then quasi-stable, flying of order centimeters before decay, leaving a highly ionizing track (HIT) which terminates upon decay into very soft decay products. Chargino pair production could be revealed
at ILC via initial state radiation of a hard photon, and then identification of one or more
HITs, or stubs. In addition, if $\sqrt{s}$ is increased to $\sim 1$ TeV, then $\tz_1\tz_2$
production opens up, although rates are expected to be small. 
In this case, one expects $\tz_2\to W\twpm_1$  or $\tz_1 h$ to occur.

\subsection{Br\"ummer-Buchm\"uller (BB) benchmark}
\label{sec:bb}

Br\"ummer and Buchm\"uller have proposed a model wherein the Fermi scale
emerges as a focus point within high scale gauge mediation~\cite{Brummer:2012zc}.
The model is inspired by GUT-scale string compactifications which frequently predict
a large number of vector-like states in incomplete GUT multiplets which may serve as
messenger fields for gauge mediated SUSY breaking which is implemented at or around the
GUT scale. By adopting models with large numbers of messenger fields, it is found that the
weak scale emerges quite naturally from the scalar potential as a focus point from
RG running of the soft terms. 
The soft SUSY breaking terms receive both gauge-mediated and gravity-mediated contributions.
The gauge-mediated contributions are dominant for most soft masses, while the $A$-terms and $\mu$ may be forbidden by symmetry.
The superpotential higgsino mass term $\mu$ emerges from gravitational interactions and
is expected to be of order the gravitino mass $\mu\sim m_{3/2}\sim 150-200$ GeV.
The spectrum which emerges from the model tends to contain gluino and squark masses in the 
several TeV range so that the model is compatible with LHC constraints.
States accessible to a linear collider would include the higgsino-like light charginos $\twpm_1$
and neutralinos $\tz_{1,2}$ similar to the Hidden SUSY model in Subsection~\ref{sec:hs}.

For ILC studies, we adopt the benchmark model with messenger indices $(N_1,\ N_2,\ N_3)=(17,23,9)$
with $\tan\beta =52$ and weak scale values of $\mu=200$ GeV and $m_A=1120$ GeV, with $A_i\simeq 0$.
Then the GUT scale scalar masses are found to be: 
$m_Q=1538.5$ GeV, $m_U=1181.2$ GeV, $m_D=1033.8$ GeV, $m_L=1274.7$ GeV and $m_E=989.5$ GeV. 
The GUT-scale gaugino masses are given by $M_1=4080$ GeV, $M_2=4600$ GeV and $M_3=1800$ GeV.
The spectrum generated from Isasugra is listed in Table~\ref{tab:bm2}.

\begin{table}[h!]
\centering
\begin{tabular}{lcccc}
\hline
\hline
{\rm mass}        & $\delta M\ttau$ & KL & BB & NMH \\
\hline
$m_h$             & 0.124           & 0.125 & 0.123 & 0.125 \\
$m_{A}$           & 0.400           & 2.0   & 1.120 & 5.32 \\
$m_{H}$           & 0.401           & 2.013 & 1.127 & 5.35 \\
$m_{H^{\pm}}$     & 0.408           & 2.014 & 1.131 & 5.36 \\
$\mu$             & 0.2             & 2.0   & 0.2  & 3.0 \\
\hline
$m_{\tilde{g}}$                & 2.0          & 1.79  & 3.817 & 1.496  \\
$m_{\tilde{\chi}^{\pm}_{1,2}}$ & 0.155, 0.282 & 0.2004, 2.05 & 0.214, 3.76 & 0.535, 3.0 \\
$m_{\tilde{\chi}^0_{1,2}}$     & 0.097, 0.162 & 0.2001, 0.616 & 0.0.205, 0.208 & 0.277, 0.533 \\
$m_{\tilde{\chi}^0_{3,4}}$     & 0.209, 0.282 & 2.05, 2.05 & 1.83, 3.78 & 2.99, 3.0 \\
\hline 
$m_{ \tilde{u}_{L,R}}$   & 2.03, 2.03  & 24.8, 25.3 & 4.55, 3.56 & 1.237, 1.215 \\
$m_{\tilde{t}_{1,2}}$    & 0.299, 1.53 & 16.4, 20.9 & 2.28, 3.85 & 1.998, 3.763 \\
\hline 
$m_{ \tilde{d}_{L,R}}$   & 2.03, 2.03  & 24.8, 24.8 & 4.55, 3.41 & 1.24, 1.167  \\
$m_{\tilde{b}_{1,2}}$    & 0.338, 1.53 & 20.8, 24.7 & 2.54, 3.85 & 3.789, 4.874  \\
\hline
$m_{ \tilde{e}_{L,R}}$   & 0.208, 0.135 & 25.3, 24.4 & 3.25, 1.79 & 0.507, 0.284 \\
$m_{\tilde{\tau}_{1,2}}$ & 0.104, 0.210 & 24.3, 25.2 & 0.69, 3.03 & 4.65, 4.85 \\
\hline
$\Omega_{\tz}^{std}h^2$  & 0.116 & 0.0025 & 0.008 & 0.07 \\
$\langle\sigma v\rangle(v\rightarrow 0)\times 10^{-25}\ [cm^3/s]$ & - & 19 & 1.9 & 0.0005 \\
$\sigma^{SI}(\tz p)\times 10^{9}$ [pb]         & - & 0.04 & 0.24 & 0.0012 \\
\hline
$a_\mu^{SUSY} \times 10^{10}$                  & 33.5 & 0.0002 & 0.51 & 23.4 \\
$BF(b\rightarrow s\gamma )\times 10^4$         & 3.3 & 3.2 & 3.2 & 3.2 \\
$BF(B_S\rightarrow \mu^+\mu^- )\times 10^9$    & 3.9 & 3.8 & 4.4 & 3.9  \\
$BF(B_u\rightarrow \tau\nu_\tau )\times 10^4$  & 1.1 & 1.3 & 1.1 & 1.3 \\
\hline
\hline
\end{tabular}
\caption{Input parameters and mass spectrum and rates for post LHC7
benchmark points $6-9$. All masses and dimensionful parameters are in TeV units. 
Entries marked ``-'' have not been calculated. All values are obtained from Isasugra apart from
$\delta M\ttau$, which have been calculated with SPheno and FeynHiggs (Higgs sector).}
\label{tab:bm2}
\end{table}

\subsection{Normal scalar mass hierarchy}
\label{sec:nmh}

Models with a normal scalar mass hierarchy ($m_0(1)\simeq m_0(2)\ll m_0(3)$)~\cite{Baer:2004xx} are motivated 
by the attempt to reconcile the $>3\sigma$ discrepancy in $(g-2)_\mu$ (which requires
rather light sub-TeV smuons) with the lack of a large discrepancy in $BF(b\to s\gamma )$, 
which seems to require third generation squarks beyond the TeV scale. The idea here
is to require a high degree of degeneracy amongst first/second generation sfermions
in order to suppress the most stringent FCNC processes, while allowing third 
generation sfermions to be highly split, since FCNC constraints from third generation
particles are relatively mild. The {\it normal mass hierarchy} follows in that
first/second generation scalars are assumed much lighter than third generation
scalars, at least at the GUT scale. Renormalization group running then lifts 
first/second generation squark masses to high values such that $m_{\tq}\sim m_{\tg}$.
However, first/second generation sleptons remain in the several hundred GeV range
since they have no strong coupling.

Here, we adopt a NMH benchmark point with separate ${\bf 5^*}$ and ${\bf 10}$ 
scalar masses as might be expected in a $SU(5)$ SUSY GUT model. 
We adopt the following parameters:
$m_5(3)\sim m_{10}(3)=5$ TeV, $m_{1/2}=0.63$ TeV, $A_0=-8.5$ TeV, $\tan\beta =20$, $\mu >0$
with $m_L(1,2)=m_D(1,2)\equiv m_5(1,2)=0.2$ TeV, and $m_Q(1,2)=m_U(1,2)=m_E(1,2)\equiv
m_{10}(1,2)=0.375$ TeV. The spectrum generated using Isasugra~7.82 with non-universal
scalar masses is listed in Table~\ref{tab:bm2}, where we find
$m_{\tz_1}\simeq 277$ GeV, $m_{\te_R}\simeq m_{\tmu_R}=284$ GeV, 
$m_{\tnu_{e,\mu L}}\simeq 300$ GeV and $m_{\te_L}\simeq m_{\tmu_L}=507$ GeV, as well as 
$m_h\simeq 125$ GeV. In the colored sector, $m_{\tg} =1.5$ TeV with $m_{\tq}\sim 1.2$ TeV,
so the model is compatible with LHC7 constraints, but may be testable at LHC8. 
The first and second generation squarks decay mainly into $\twpm_1 +$~jet, followed by
$\twpm_1 \ra \tnu_l l \ra \tz_1 \nu_l l$, or alternatively into $\tz_2 +$~jet, followed by
$\tz_2 \ra \tnu_l \nu_l \ra \tz_1 \nu_l \nu_l$. Thus, squark pair production will give only 2 jets,
either accompanied by just missing transverse energy or by 1 or 2 leptons. The gluino decays mostly into first or 
second generation squarks plus an additional jet. Since the $\tz_2$ decays invisibly, the only sign of direct
$\twpm_1 \tz_2$ production will be a single lepton from the the $\twpm_1$ decay plus missing transverse energy.

The model does indeed reconcile $(g-2)_\mu$ with
$BF(b\to s\gamma )$ since $\Delta a_\mu^{SUSY}\sim 23\times 10^{-10}$ and
$BF(b\to s\gamma )=3.22\times 10^{-4}$. Also, the thermal neutralino
abundance is given as $\Omega_{\tz_1}h^2\simeq 0.07$ due to neutralino-slepton
co-annihilation. An ILC with $\sqrt{s}\agt 600$ GeV would be needed to 
access the $\te_R\bar{\te}_R$ and $\tmu_R\bar{\tmu}_R$ pair production.
These reactions would give rise to very low energy di-electron and di-muon
final states which would be challenging to extract from two-photon backgrounds.
However, since it has been demonstrated that mass differences of this size are 
manageable even in the case of $\tau$ leptons from $\ttau$ decays~\cite{Bechtle:2009em}, 
it should be feasible also in case of electrons or muons.
Since $\tnu\to\nu+\tz_1$, sneutrinos would decay invisibly, although the reaction
$e^+e^-\to\tnu_L\bar{\tnu}_L\gamma$ may be a possibility. The lack of
$\ttau^+\ttau^-$ pair production might give a hint that nature is described by 
a NMH model.

\section{Conclusions}
\label{sec:conclude}

At first sight, it may appear very disconcerting that after one full
year of data taking at LHC7, with $\sim 5$ fb$^{-1}$ per experiment, no sign of supersymmetry
is yet in sight. On the other hand, evidence at the $3\sigma$ level seems to be emerging that
hints at the presence of a light higgs scalar with mass $m_h\sim 125$ GeV. While 
$m_h$ can theoretically inhabit a rather large range of values of up to $800$~GeV in the Standard Model, 
the simplest supersymmetric extensions of the SM require it to lie below $\sim 135$ GeV.
A light SUSY Higgs of mass $\sim 125$ GeV seems to require top squark masses $m_{\tst_i}\agt 1$ TeV
with large mixing: thus, the emerging signal seems more consistent with a super-TeV sparticle mass
spectrum than with a sub-TeV spectrum, and indeed the latter seems to be nearly excluded by
LHC searches for gluinos and first and second generation squarks (unless there is a highly compressed 
spectrum, or other anomalies).
In addition, a Higgs signal around $125$~GeV highly stresses at least the minimal versions of
constrained models such as AMSB and GMSB, and may favor gravity-mediated SUSY breaking models which
naturally accommodate large mixing in the top squark sector.

While some groups had predicted just prior to LHC running a very light sparticle mass spectrum
(based on global fits of SUSY to a variety of data, which may have been overly skewed by the
$(g-2)_\mu$ anomaly), the presence of a multi-TeV spectrum of at least first/second generation
matter scalars was not unanticipated by many theorists. The basis of this latter statement
rests on the fact that a decoupling of first/second generation matter scalars either solves or
at least greatly ameliorates: the SUSY flavor problem, the SUSY $CP$ problem, the SUSY GUT proton decay
problem and, in the context of gravity mediation where the gravitino mass sets the scale for
the most massive SUSY particles, the gravitino problem. 

In contrast, examination of electroweak fine-tuning arguments, applied to the radiatively corrected
SUSY scalar potential imply that models with 1. low $|\mu |\alt \Lambda_{NS}\sim 200$ GeV, 
2. third generation squarks with $m_{\tst_{1,2},\tb_1}\alt 1.5$ TeV and 3. $m_{\tg}\alt 4 $ TeV
are favored. Since first/second generation matter scalars don't enter the electroweak
scalar potential, these sparticles can indeed exist in the 10-50 TeV regime -- as required by decoupling --
without affecting fine-tuning. The class of models which fulfill these conditions are called 
{\it natural SUSY} or NS models. NS models are typically very hard to detect at LHC unless some
third generation squarks are very light $\sim 200-600$ GeV, with a large enough decay mass gap 
to yield sufficient visible energy. The set of light higgsinos $\twpm_1$, $\tz_2$ and $\tz_1$
can be produced at high rates at LHC, but the very tiny visible energy release from higgsino decays
makes them exceedingly hard to detect. However, NS at an ILC may well be a boon! An ILC would
likely then be a {\it higgsino factory} in addition to a Higgs factory. The small visible
energy release from higgsino-like chargino decays should be visible against backgrounds originating from
two-photon initiated processes, especially when an additional hard ISR photon is required. 
In addition, there is a good chance that some or even most third generation squarks and 
sleptons may be accessible given high enough beam energy. 
As the fine-tuning upper bound $\Lambda_{NS}$ increases, the NS spectrum blends into Hidden SUSY where 
the higgsinos are still light, but the third generation is lifted beyond LHC/ILC reach. 
The HS collider phenomenology is expected to be very similar to that emerging from a 
non-minimal GMSB model suggested by Br\"ummer and Buchm\"uller (BB).

We also present several benchmark models consistent with LHC and other constraints which predict some varied 
phenomenology. One NUHM2 point with heavy matter scalars and $m_h=125$ GeV contains $A$ and $H$ Higgs bosons
which would also be accessible to ILC. A model with non-universal gaugino masses (NUGM) allows for
chargino pair production at ILC followed by $\twpm_1\to W\tz_1$ decay, leading to $W^+W^- +\esl$ events.
Also, a rare surviving benchmark from mSUGRA/CMSSM is presented in the far focus point region
with $m_h=125$ GeV, with matter scalars at $m_{\tq,\tell}\sim 15$ TeV, 
where chargino pairs of the mixed bino-higgsino variety are accessible to an ILC.
We also present one benchmark point from the Kallosh-Linde (KL) model. In this case, matter scalars have masses
$m_{\tq,\tell}\sim m_{3/2}\sim 25$ TeV, but gaugino masses follow the AMSB pattern, with the $\twpm_1$ and $\tz_1$
being nearly pure wino, with $m_{\twpm_1}-m_{\tz_1}\sim 0.33$ GeV mass gap. If the mass gap is small enough, then 
charginos can fly a measureable distance before decay. It might be possible to detect 
$e^+e^-\to \twp_1 \twm_1 \gamma \to\gamma+$ soft debris including possible highly ionizing tracks which
terminate into soft pions. The phenomenology of this model is similar to that expected from G2MSSM of
Acharya {\it et al.}~\cite{Acharya:2008zi}.
Finally, we present pMSSM and NMH models with light charginos and sleptons which is in accord with the 
$(g-2)_\mu$ anomaly, $m_h\sim 124$ GeV and with a standard neutralino relic abundance 
$\Omega_{\tz_1}^{std}h^2=0.11$. The ILC-relevant part of the spectrum is very similar to the well-studied 
SPS1a scenario~\cite{Allanach:2002nj} (or its variant SPS1a').

In summary, results from the LHC7 run in 2011 have resulted so far in no sign of SUSY particles, 
although impressive new limits on gluino and squark masses have been determined. 
In addition, much of the 
expected mass range for a SM-like Higgs boson has been ruled out save for the narrow window
of $115$ GeV $<M_H<$ 127 GeV. Indeed, within this window, there exists $\sim 3\sigma$ hint
for a 125 GeV Higgs signal in several different channels from both Atlas and CMS, and also from
CDF/D0 at the Fermilab Tevatron. If the Higgs-hint is verified, this can be regarded as an overall
positive for weak scale supersymmetry in that the Higgs would fall squarely within the
narrow predicted SUSY window. While the lack of gluino and first generation squark signals at LHC7 may at
first be disconcerting, it must be remembered that first generation squarks, and to some degree
gluinos, contribute little to naturalness arguments which connect SUSY breaking to the weak scale.
Naturalness arguments do favor a value of $\mu\sim M_Z$, with perhaps $\mu$ ranging as high as
$\sim 200$ GeV. In this case, a spectrum of light higgsinos is anticipated. Such light higgsinos
would be very difficult to detect at LHC, while an ILC with $\sqrt{s}=0.25-1$ TeV would
be a {\it higgsino factory}, in addition to a Higgs factory! 
Naturalness arguments, and also the
muon $g-2$ anomaly, portend a rich assortment of new matter states likely accessible to the ILC,
although such states will be difficult for LHC to detect. We hope the benchmark models
listed here give some view as to the sort of new SUSY physics which may be
expected at ILC in the post LHC7 era.

\section{Acknowledgments}

We thank Mikael Berggren, Azar Mustafayev, Krzysztof Rolbiecki and Annika Vauth for 
supporting calculations and valuable discussions, and Benno List and Xerxes Tata for comments on the manuscript.

\section{Bibliography}


\begin{footnotesize}


\end{footnotesize}


\end{document}